\documentclass[twocolumn, trackchanges]{aastex63}

\usepackage{multirow}
\usepackage{asmath}
\usepackage{graphicx}
\usepackage{bm}
\usepackage[T1]{fontenc}

\newcommand{\myemail}{cornachi@usna.edu}
\newcommand{\Romannumeral}[1]{\uppercase\expandafter{\romannumeral #1\relax}}

\shorttitle{Shallow Quasar Accretion Disk Profiles}
\shortauthors{Cornachione et al.}

\published{June 1, 2020}

\begin{document}

\title{Quasar Microlensing Variability Studies Favor Shallow Accretion Disk Temperature Profiles.}

\author[0000-0003-1012-4771]{Matthew A. Cornachione}
\affiliation{Department of Physics, United States Naval Academy, 572C Holloway Rd., Annapolis, MD 21402, USA}
\email{\myemail}

\author[0000-0003-2460-9999]{Christopher W. Morgan}
\affiliation{Department of Physics, United States Naval Academy, 572C Holloway Rd., Annapolis, MD 21402, USA}






\begin{abstract}
We compare the microlensing-based continuum emission region size measurements in a sample of 15 gravitationally lensed quasars with estimates of luminosity-based thin disk sizes to constrain the temperature profile of the quasar continuum accretion region.  If we adopt the standard thin disk model, we find a significant discrepancy between sizes estimated using the luminosity and those measured by microlensing of $\log(r_{L}/r_{\mu})=-0.57\pm0.08\,\text{dex}$.  
If quasar continuum sources are simple, optically thick accretion disks with a generalized temperature profile $T(r) \propto r^{-\beta}$, the discrepancy between the microlensing measurements and the luminosity-based size estimates can be resolved by a temperature profile slope $0.37 < \beta < 0.56$ at $1\,\sigma$ confidence.
This is shallower than the standard thin disk model ($\beta=0.75$) at $3\,\sigma$ significance.    We consider alternate accretion disk models that could produce such a temperature profile and reproduce the empirical continuum size scaling with black hole mass, including disk winds or disks with non-blackbody atmospheres.
\end{abstract}

\section{Introduction}
\label{sec:intro}

Quasars play a critical role in galaxy evolution and growth \citep{boyl1998a, dima2005a}. Because of the massive size of their central black holes, they may also serve as tests of general relativity \citep{fabi2000a, char2017a, dai2019a}.  Despite the astrophysical importance of quasars and significant effort in the field, we still have an incomplete picture of the physical processes governing quasar emission \citep[e.g.][]{blae2004a, anto2013a}.  No model of quasar accretion disks can fully explain the wide range of observed phenomena, including spectral slope scaling with mass \citep{davi2007a}, the ${\sim}1000\,\text{\AA}$ spectral cutoff \citep{shul2012a}, and the large optical/UV continuum emission region size \citep{morg2010a}.

Despite some limitations, the thin-disk model of \citet[SS73]{shak1973a} and general relativistic extensions \citep[e.g.][]{novi1973a} have provided a coherent framework within which to interpret observed quasar behavior.  In the thin disk model, a hot disk accretes onto the central black hole, with energy transported outward through viscous dissipation.  
The SS73 thin disk model remains in wide use in the present day mostly due to its relatively simple analytic form.


Subsequent studies have developed progressively more realistic quasar accretion models. \citet{abra1988a} proposed a `slim disk' model in which vertical energy transfer within the disk plays an important role. This changes the disk dynamics and leads to advective flows radially within the disk when accretion rates rise above roughly $\dot{m} \equiv \dot{M}_{\text{BH}}/(16\dot{M}_{\text{Edd}}) = 0.1$.  Although this model and more recent extensions require numerical solutions, they may better describe the soft X-ray spectra in high-luminosity AGN and can explain long optical-UV lags \citep{szus1996a, sado2011a}.
A key advance came when \citet{balb1991a} demonstrated that a small magnetic field can create turbulence in a disk.  These magnetorotational instabilities (MRI) finally gave a physically plausible source of the accretion disk viscosity required in the thin and slim disk models.  Since this discovery, magnetic field contributions are included in many accretion disk models \citep[e.g.,][]{agol2000a, penn2010a, sado2016a, jian2019a}. State-of-the-art models typically use numerical general relativistic magnetohydrodynamic (GRMHD) simulations to predict accretion disk properties \citep[see][]{port2019a}.  Most of these simulations are, however, presently limited to low accretion rates rather than conditions found in typical quasars.


One of the well-known shortcomings of the thin disk model is its inability to correctly predict the size of accretion disks from their observed luminosities.  Assuming blackbody emission within each annulus of the disk, we can estimate the size of an optically thick accretion disk using the observed specific flux, reported as the half light radius $r_{L}$,
\begin{align}
\begin{split}
r_{L} = C(\beta)r_{\lambda} = C(\beta)&\left(\frac{\lambda_{obs}^5D_{L}^2F_{\lambda, obs}}{4\pi h c^2 \cos(i) (1+z)^4}\right)^{1/2}\times \\ &\left( \int_{u_{\text{in}}}^\infty \frac{u\,du}{\exp{\left(u^{\beta}\right)}-1}. \right)^{-1/2}.
\label{eqn:rlambda_flux}
\end{split}
\end{align}
Here $F_{\lambda, obs}$ is the observed flux per wavelength in a particular band centered at the observed wavelength $\lambda_{obs}$ and $D_{L}$ is the luminosity distance to the quasar.  The size $r_{\lambda}$ is the scale radius of the disk at this wavelength, defined such that $kT(r_{\lambda_{\text{rest}}}) = hc/\lambda_{\text{rest}}$ (see Equation 2 of \citet{morg2010a} or Equation 5 of \citet{li2019a}).  The integral is over the dimensionless parameter $u=r/r_{\lambda}$.
The factor of $\cos(i)$ accounts for disk inclination.  For later generalization, we leave this as a function of the temperature profile slope $\beta$, where $\beta=3/4$ for the \citet{shak1973a} thin disk model.  The factor $C(\beta)$ is given in \citet{li2019a} as the numerical solution to
\begin{equation}
\int_{u_{\text{in}}}^{C(\beta)} \frac{u\,du}{(\exp{(u^{\beta})} - 1)} = \frac{1}{2} \int_{u_{\text{in}}}^\infty \frac{u\,du}{(\exp{(u^{\beta})} - 1)}
\label{eqn:cbeta}
\end{equation}
and converts the scale radius to a half light radius.  To a good approximation, we can set $u_{\text{in}}=0$ for optical and infrared $\lambda_{obs}$.  For a standard thin disk this gives $C(3/4)=2.44$.  

Equation~\ref{eqn:rlambda_flux} provides estimates of the quasar size that are systematically smaller by a factor of ${\sim}3\text{--}4$ than the measurements from microlensing \citep{pool2007a, morg2010a, blac2011a} and reverberation mapping \citep{cack2007a,  bent2010a, edel2015a}.  A number of groups have recently posited a variety of new models, some of which might resolve this discrepancy.  \citet{dext2011a} demonstrated that an inhomogeneous disk could reproduce the correct disk size, although the level of inhomogeneity required may not be physically plausible.  The disk structure may be different at large radii \citep{jian2016a}, leading to an adjustment in inferred luminosity sizes.  A modified disk atmosphere, such as that proposed in \citet{hall2018a} would give rise to a non-blackbody spectrum which could also yield luminosity-based sizes consistent with microlensing sizes.  Alternatively, the disk could have mass outflow in the form of disk winds \citep[e.g.][]{prog2005a, slon2012a, yuan2012a, laor2014a, tomb2015a, li2019a}, which could also match measured disk sizes.  More complex physics, included in GRMHD simulations, can potentially also reproduce measured continuum emission region sizes.

A more sophisticated picture comes from constraining the shape of the disk continuum emission through modeling or observation.  Keeping for now the assumption of blackbody emission within each annulus, and examining regions far from the inner disk edge, the effective surface temperature profile takes on the relatively simple form of
\begin{equation}
T(r) \propto r^{-\beta}
\label{eqn:t_v_beta}
\end{equation}
where $T$ is the disk temperature, $r$ is the radius from the center of the black hole, and $\beta$ is the temperature profile slope.  The thin disk model predicts a temperature profile slope of $\beta=3/4$.  The slim disk model of \citep{abra1988a} predicts a shallower temperature profile slope of $\beta\approx0.5$ for high accretion rates (estimated from Figure 11 in \citet{abra1988a}).   Other models include a non-zero torque on the inner disk boundary \citep{agol2000a,penn2012a}, and the MRI model of \citet{agol2000a} explicitly predicts a steeper value of $\beta=7/8$.  \citet{sinc1998a} account for the structure of a dissipative accretion disk and found that, although the ultraviolet spectrum is not a local blackbody, the spectral slope can be nearly consistent with the thin disk model.  Other numerical models explore the impact of a radially varying disk viscosity \citep{penn2013a} and MHD spinning black holes \citep{schn2016a} though neither reports the effective temperature profile slope.

Similarly, attempts to empirically measure or estimate accretion disk temperature profiles using observational constraints have led to very inconsistent results. 
In the rest-frame optical and ultraviolet, \citet{vand2001a} and \citet{gask2008a} estimated a temperature profile slope of $\beta=0.57$ using the spectral continuum slope \citep{sun1989a}.  A comparably shallow temperature profile slope is inferred in several similar studies \citep{davi2007a, bonn2013a, xie2016a}, but the authors emphasize that dust extinction in the host galaxy can significantly affect the observed temperature profile.  This host extinction is difficult to correct for and gives rise to high uncertainty in the intrinsic temperature profile slope.

Reverberation mapping (RM) studies have measured the disk temperature profile through continuum interband lags \citep[e.g.,][]{cack2007a, jian2017a, cack2018a, edel2019a}.  These studies find profile slopes consistent with the thin-disk model, although \citet{edel2019a} demonstrate that a wide range of profile slopes provide adequate fits to the observed lags.  Furthermore \citet{cack2018a} find that the broad line region can contribute to the interband lags, requiring a careful treatment of different AGN components during reverberation mapping analysis.  Although the results are consistent with the thin-disk model, reverberation mapping does not presently exclude shallower or steeper temperature profile slopes.

Temperature profile measurements also arise naturally from single-epoch microlensing measurements.  Chromatic flux ratio anomalies can lead to estimates of the quasar accretion disk size, which scales as $r\propto\lambda^{1/\beta}$.
The findings from different studies have not, however, converged to a single value.  Some studies find evidence for a shallower profile than the thin-disk model \citep{roja2014a, bate2018a} though more studies point to a steeper profile \citep{blac2011a, muno2011a, mott2012a, jime2014a, blac2015a, muno2016a, mott2017a}.  Still other studies based on multi-epoch, multi-filter curves conclude that a thin-disk model can explain observations \citep[e.g.,][]{eige2008a, mosq2011b}.  Studies such as that by \citet{bate2018a} have begun to explore the systematic uncertainties that have lead to such a wide spread in measurements.  They demonstrated that in systems with low chromatic variation, the single epoch method will return an overly steep profile, and they propose a means to quantify this effect in future work.

\citet{morg2010a} developed an alternate, model dependent, means to constrain the temperature profile using the observed difference between the accretion disk size measured by microlensing, $r_{\mu}$, \citep[e.g.][]{koch2004a, bate2008a} and that inferred from the quasar luminosity, $r_{L}$.  By modeling the expected ratio $r_{L}/r_{\mu}$ as a function of temperature profile slope, constraints on $\beta$ can be visualized.  In \citet{morg2010a} the authors found evidence for a shallower-than-thin-disk temperature profile, but they did not exclude thin disk models. With a larger sample size, we are now able to employ this technique to tightly constrain the slope of the accretion disk temperature profile.  

This paper is organized as follows.  In Section~\ref{sec:data}  we present the discrepancy between microlensing and luminosity-based continuum sizes in our sample of quasars with multi-epoch microlensing lightcurves.  In Section~\ref{sec:results} we show how microlensing and luminosity sizes might be brought into agreement with a different temperature profile slope $\beta$.  We also explore the impact of contaminating light sources.  We compare our results to alternate disk models, focusing on two recent analytic models from \citet{hall2018a} and \citet{li2019a}. We summarize our conclusions and discuss future observational programs in Section~\ref{sec:discussion}.  In all calculations we adopt a flat cosmological model with $\Omega_M = 0.3$, $\Omega_{\Lambda} = 0.7$, and $H_0 = 70\,\text{km}\,\text{s}^{-1}\,\text{Mpc}^{-1}$ \citep{hins2009a}.

\section{Data}
\label{sec:data}

\begin{figure}[t]
\centering
\includegraphics[width=0.5\textwidth]{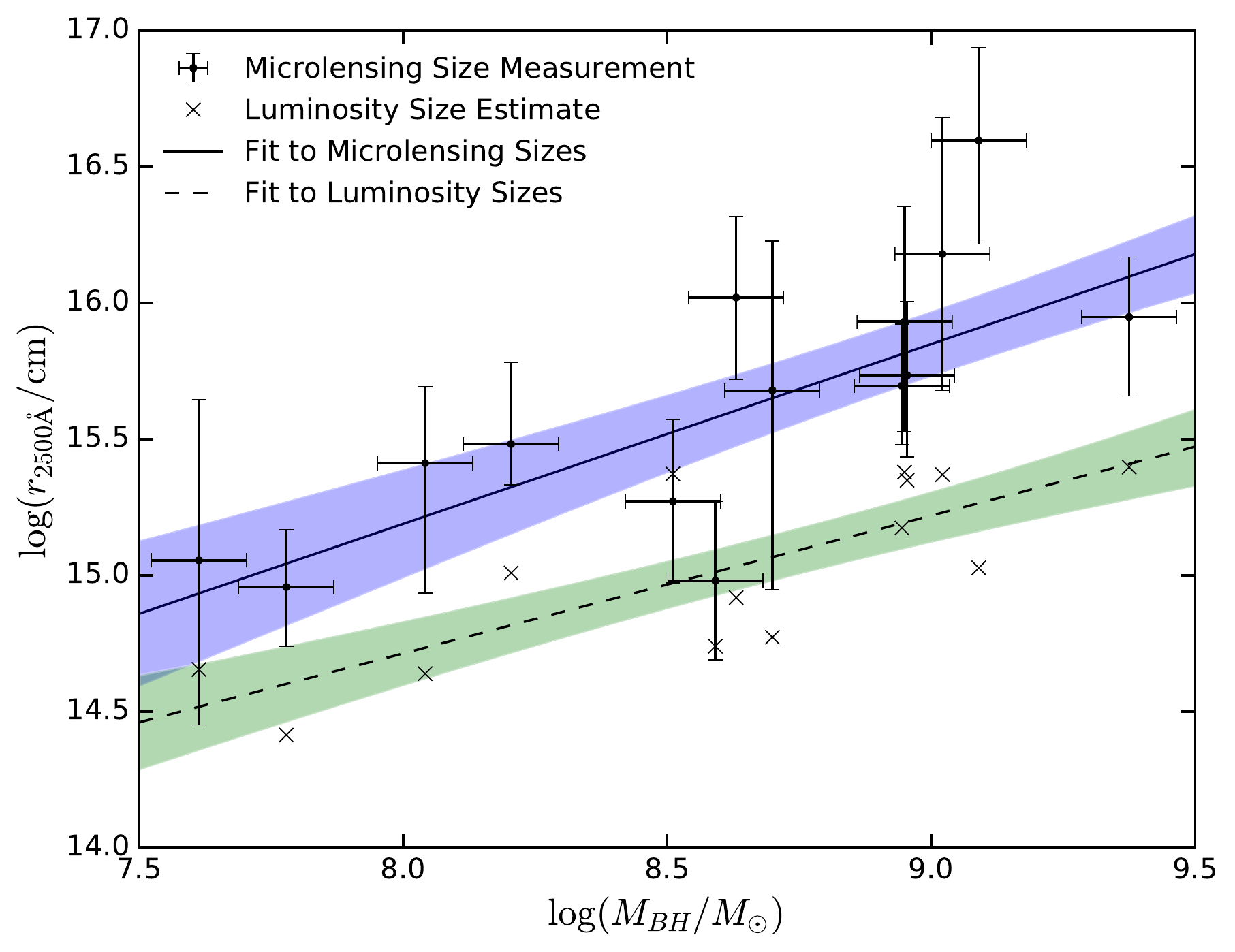} 
\caption{Quasar Accretion Disk Size-Black Hole Mass Relation.  The accretion disk sizes are scaled to $\lambda_{\text{rest}} = 2500\text{\AA}$ \citep{koch2004a, morg2008a, dai2010a, morg2010a, hain2012a, morg2012a, hain2013a, macl2015a,morg2018a,corn2020a}.  Microlensing sizes are shown as black dots with error bars and luminosity size estimates are shown as crosses (errors omitted).  The best-fit lines are shown along with the $1\sigma$ errorbars for the microlensing size measurements (blue, solid) and the luminosity size estimates (green, dashed).}
\label{fig:r_v_mbh}
\end{figure}

\begin{figure}[t]
\centering
\includegraphics[width=0.5\textwidth]{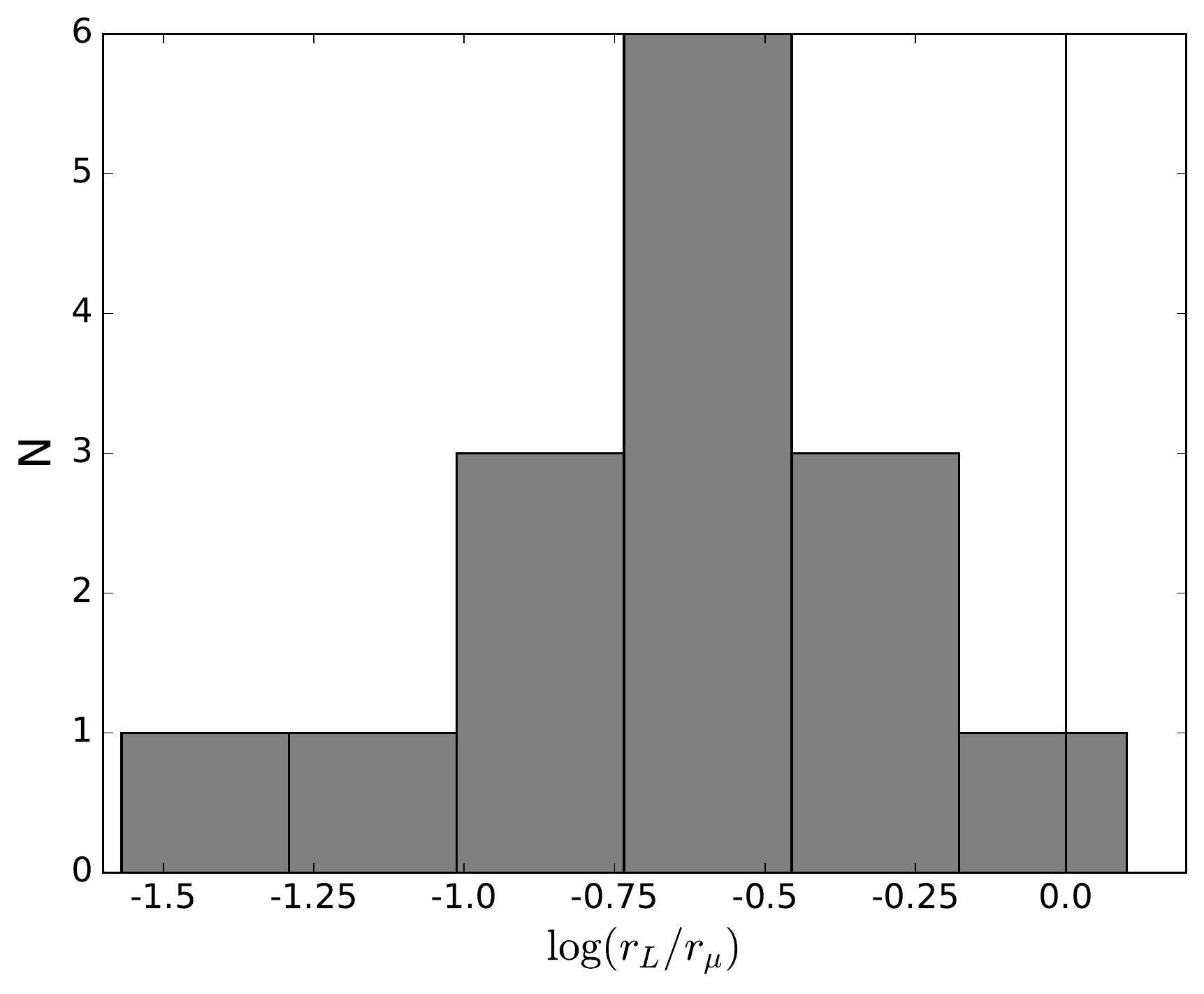} 
\caption{Histogram of the luminosity and microlensing size ratio, $\log(r_{L}/r_{\mu})$, for the 15 systems in our sample.  The distribution peaks near the average of -0.57 with an error-weighted standard deviation of ${\sim}0.08$.}
\label{fig:r_ratio_hist}
\end{figure}

\begin{deluxetable}{lccc}
\tablecaption{Microlensing size measurements of the continuum size $\log(r_{\mu})$,  luminosity continuum size estimates for a standard thin disk ($\beta=3/4$), $\log(r_{L})$, and the ratio, $\log(r_{L}/r_{\mu})$, for our sample of 15 lensed quasars.  All sizes are scaled to $2500\,\text{\AA}$ in the quasar rest frame and a nominal inclination angle of $60\,\degr$.  Microlensing size errors are found from the $16^{\text{th}}$ and $84^{\text{th}}$ percentile values of the posterior size distributions.  Luminosity size uncertainty is calculated from the intrinsic flux uncertainty, which includes both photometric error and magnification uncertainty. \label{tab:qso_stats}}
\tablehead{
\colhead{Quasar} & \colhead{$\log(r_{\mu}/\text{cm})$} & \colhead{$\log(r_{L}/\text{cm})$} & \colhead{$\log(r_{L}/r_{\mu})$}
}
\startdata
QJ0158 & $ 15.87^{+  0.30}_{-  0.15}$ & $ 15.40\pm  0.05$ & $  -0.47\pm  0.23$ \\
HE0435 & $ 16.07^{+  0.55}_{-  0.73}$ & $ 15.16\pm  0.10$ & $  -0.91\pm  0.65$ \\
SDSS0924 & $ 15.80^{+  0.28}_{-  0.48}$ & $ 15.03\pm  0.10$ & $  -0.77\pm  0.39$ \\
FBQ0951 & $ 16.32^{+  0.42}_{-  0.40}$ & $ 15.77\pm  0.04$ & $  -0.55\pm  0.42$ \\
SDSS1004 & $ 15.37^{+  0.29}_{-  0.29}$ & $ 15.13\pm  0.18$ & $  -0.24\pm  0.34$ \\
HE1104 & $ 16.34^{+  0.22}_{-  0.29}$ & $ 15.79\pm  0.13$ & $  -0.55\pm  0.28$ \\
PG1115 & $ 16.99^{+  0.34}_{-  0.38}$ & $ 15.42\pm  0.11$ & $  -1.57\pm  0.38$ \\
RXJ1131 & $ 15.35^{+  0.21}_{-  0.22}$ & $ 14.80\pm  0.04$ & $  -0.54\pm  0.22$ \\
SDSS1138 & $ 15.44^{+  0.59}_{-  0.60}$ & $ 15.04\pm  0.08$ & $  -0.40\pm  0.60$ \\
SBS1520 & $ 16.08^{+  0.23}_{-  0.22}$ & $ 15.56\pm  0.05$ & $  -0.52\pm  0.23$ \\
Q2237 & $ 16.12^{+  0.27}_{-  0.30}$ & $ 15.74\pm  0.18$ & $  -0.39\pm  0.34$ \\
WFI2033 & $ 16.41^{+  0.30}_{-  0.30}$ & $ 15.31\pm  0.10$ & $  -1.10\pm  0.32$ \\
SBS0909 & $ 15.66^{+  0.30}_{-  0.30}$ & $ 15.76\pm  0.04$ & $ 0.10\pm  0.30$ \\
Q0957 & $ 16.57^{+  0.50}_{-  0.50}$ & $ 15.76\pm  0.03$ & $  -0.81\pm  0.50$ \\
WFI2026 & $ 16.13^{+  0.29}_{-  0.34}$ & $ 15.47\pm  0.12$ & $  -0.66\pm  0.34$ \\
\enddata
\end{deluxetable}

Accretion disk microlensing size measurements are consistent between the single and multi-epoch approaches  \citep[e.g.][]{pool2007a, morg2010a, blac2011a}. The multi-epoch Monte Carlo lightcurve analysis technique of \citet{koch2004a} is, however, effectively independent of the most sensitive statistical prior, the mean mass of a microlens star in the lens galaxy.  In this technique, the quasar is modeled as a disk of varying size crossing a range of possible stellar magnification fields.  By coupling these results with cosmological velocity models, one does not need to assume a mean microlens star mass.
 $\,$We therefore restrict our analysis to only those systems with microlensing sizes measured from multi-epoch lightcurves, given in Table~\ref{tab:qso_stats}.  These values and uncertainties are found from the median and $16^{\text{th}}$ to $84^{\text{th}}$ percentile range of the the reported posterior microlensing size distributions.  These 15 systems are shown in Figure~\ref{fig:r_v_mbh} with both microlensing and luminosity sizes as a function of black hole mass, where the size offset is readily apparent \citep{koch2004a, morg2008a, dai2010a, morg2010a, hain2012a, morg2012a, hain2013a, macl2015a,morg2018a,corn2020a}.

In lensed quasars we observe individual image fluxes.  Using these and magnifications from strong lensing models we calculate a quasar's intrinsic luminosity.  To estimate the magnification, most of these studies generated a range of galaxy mass models with varying smooth matter fractions (see for example Section~4.1 of \citet{morg2018a}).  These sequences give both the typical magnification and the magnification uncertainty range.
The estimated lens magnification uncertainty usually dominates the photometric measurement error, leaving the typical intrinsic flux uncertainty on the order of  $0.1\,\text{dex}$.  Adopting these flux values and uncertainties in each system and, for the moment, assuming a standard thin disk model,  we can find the luminosity-based continuum emission region sizes, reported as a half light radius $r_{L}$, using Equation~\ref{eqn:rlambda_flux}.  These luminosity-based size estimates are given in Table~\ref{tab:qso_stats}, along with the ratio $\log(r_{L}/r_{\mu})$.  Since $r_{L}$ and $r_{\mu}$ measure the same physical quantity, this ratio must be unity for any valid disk model.

Of these 15 quasars, the thin disk luminosity sizes are smaller than the microlensing sizes in all but one system as seen in Figure~\ref{fig:r_ratio_hist}.
Only SBS 0909+532 has a luminosity size larger than the microlensing size, but the intrinsic luminosity of this system and luminosity-based size estimate are uncertain because of significant systematic differences between the lens models and magnifications predicted by the astrometric fits in \citet{leha2000a} and \citet{slus2012a}.  Nevertheless, we include this system for completeness.
$\,$The error-weighted mean offset between the luminosity and microlensing size, $\log(r_{L}/r_{\mu})$, is $-0.57\pm0.08\,\text{dex}$, assuming standard thin disk theory ($\beta=3/4$).   We use this ratio as a constraint in subsequent modeling. 

The ratio $\log(r_{L}/r_{\mu})$ shows no significant dependence on black hole mass.  We do find, however, that the measured microlensing sizes scale with black mass as $r_{2500\text{\AA}} \propto M_{BH}^{0.66\pm0.15}$ \citep{morg2018a}.  This scaling is nearly identical to the thin disk prediction of $r \propto M_{BH}^{2/3}$, a curious result given the clear inconsistency between the thin disk luminosity sizes and microlensing measurements.  This empirical scaling relation with black hole mass must still be reproduced by any alternate disk model, a point we explore in Section~\ref{sec:size_v_mbh}.

\section{Results}
\label{sec:results}

\begin{figure}[t]
\centering
\includegraphics[width=0.5\textwidth]{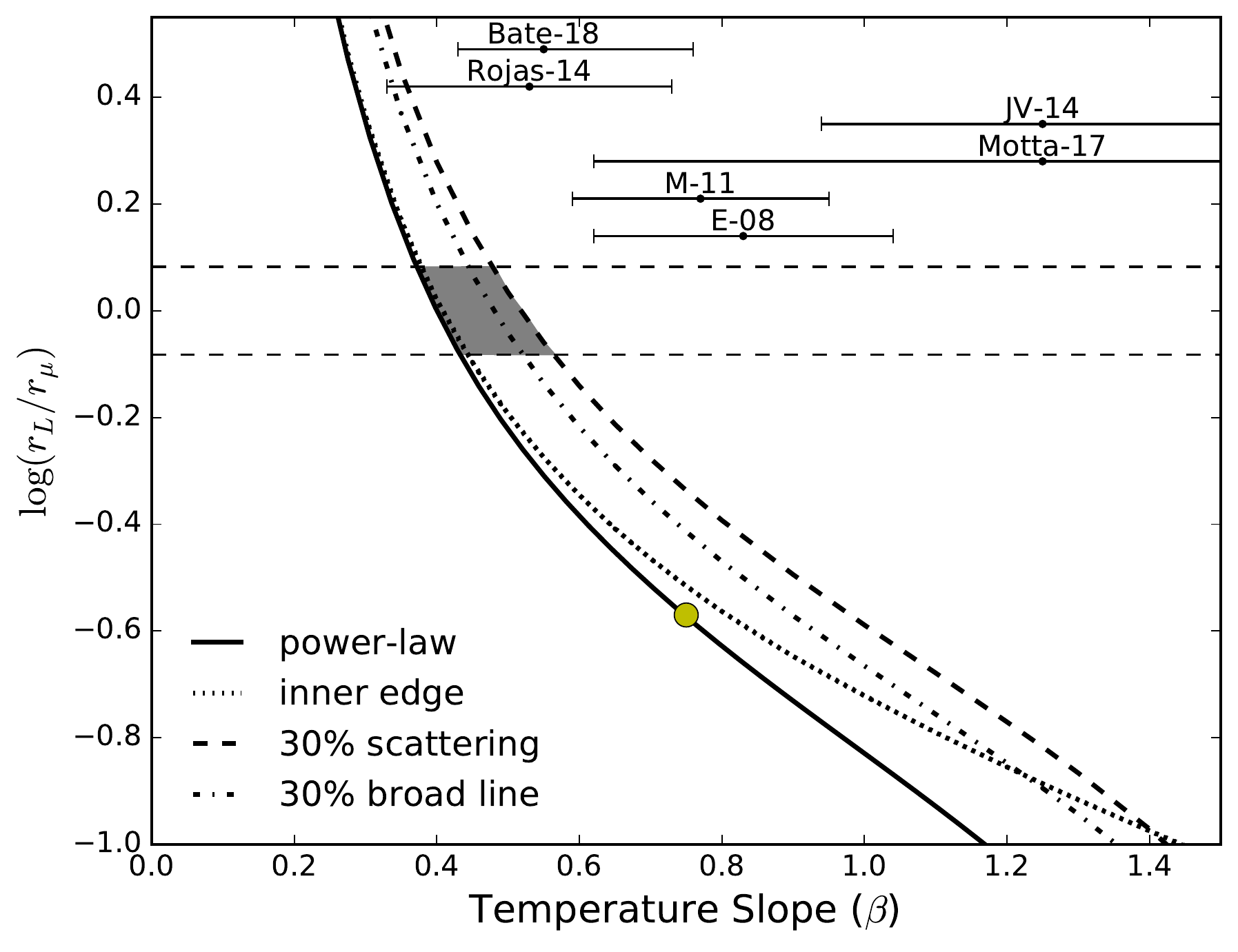}
\caption{Variation in the size ratio, $\log(r_{L}/r_{\mu})$, between the luminosity half light radius, $r_{L}$, and the microlensing half light radius, $r_{\mu}$, as a function of $\beta$.  The yellow circle indicates the ratio found from the thin disk luminosity size estimate, $\log(r_{L}/r_{\mu}) = -0.57$ at $\beta=3/4$.  If luminosity-based sizes are to be brought into agreement with microlensing sizes, the ratio $r_L/r_{\mu}$ must fall within the region defined by the dashed horizontal lines $\log(r_L/r_{\mu}) = 0 \pm 0.08$.  The solid curve shows sizes estimated with our generalized temperature profile, fixing the inner radius to zero ($r_{\text{in}}=0$).  The dotted curve shows the impact of adding a finite inner disk radius at $r_{\text{in}}=0.1r_{\lambda}$, which only results in significant differences at high $\beta$.  The effects of contamination due to either scattering (dashed) or the broad-line region (dot-dashed) are also shown.  The shaded region represents the likely parameter space $0.37 < \beta < 0.56$ under this range of models.    We also indicate temperature profile estimates from other microlensing studies \citep{eige2008a,mosq2011a, jime2014a, roja2014a, mott2017a, bate2018a}}
\label{fig:tslope}
\end{figure}

In this work we explore several possible solutions to the size inconsistency described in Section~\ref{sec:data}.  First we quantify the ratio offset, $\log(r_{L}/r_{\mu})$, as a function of $\beta$,  adopting the effective surface temperature profile in Equation~\ref{eqn:t_v_beta}, $T(r)\propto r^{-\beta}$.  We calculate the dependence of the luminosity continuum sizes on the temperature profile slope.  While microlensing size measurements are effectively independent of continuum region shape, we consider the impact on these measurements due to contaminating flux from large scale emission.  We also recognize that quasar accretion disks may not be homogenous, optically thick emitters, in which case the generalized temperature profile slope of Equation~\ref{eqn:t_v_beta} need not describe the surface temperature.  We depart from the simplest assumptions and consider alternate disk models that could reconcile microlensing size measurements with observed fluxes.  Finally we show that the empirical scaling of continuum emission size with black hole mass provides another test for any proposed disk model.

\subsection{Size Ratio vs. Temperature Profile Slope}

\begin{figure}[t]
\centering
\includegraphics[width=0.5\textwidth]{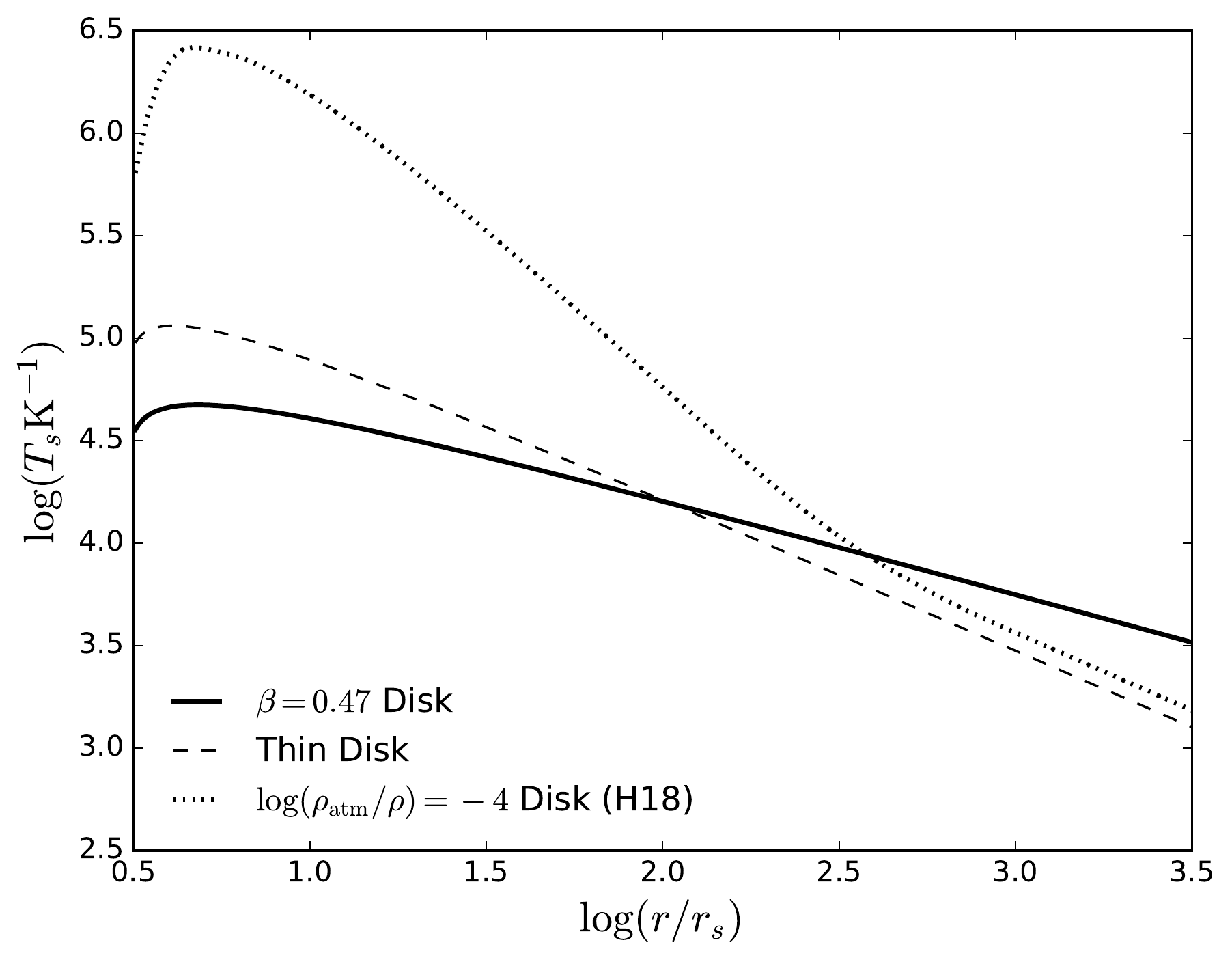}
\caption{Disk surface temperature, $T_s$, as a function of radius shown for a black hole of $10^8\,M_{\odot}$.  Radii are scaled to the Schwarzchild radius, $r_s$, with an inner edge set to $3r_s$.  Our shallow disk profile, with normalization fixed by our size measurements, is shown in solid black for our nominal value of $\beta=0.47$.  The standard thin disk is shown with a dashed black line.  A disk atmosphere model from \citet{hall2018a} with a characteristic constant atmosphere of $\log(\rho_{\text{atm}}/\rho)=-4$ is included as a dotted black line.}
\label{fig:temp_v_r}
\end{figure}

Based on the findings of \citet{mort2005a}, size, not shape, dominates microlensing variability.  This means that the microlensing continuum size measurement, $r_{\mu}$, is effectively independent of $\beta$.  Although \citet{mort2005a} and \citet{cong2007a} do find a small difference between microlensed images due to shape, any changes in $r_{\mu}$ with $\beta$ will have a minor influence on the observed size ratio.
The dominant $\beta$ dependence, therefore, comes from the luminosity size estimate, $r_{L}$.

The luminosity continuum size estimate, however, varies significantly with changes in $\beta$.  We use luminosity sizes from Equation~\ref{eqn:rlambda_flux} and numerically calculate $\log({r_{L}}/{r_{\mu}})$ vs. $\beta$.  We plot the resulting curve as a solid line in Figure~\ref{fig:tslope}, fixing the inner disk edge to $r_{\text{in}}=0$.  The yellow circle indicates the constraint from Section~\ref{sec:data} of $\log({r_{L}}/{r_{\mu}}) = -0.57\pm0.08$ at $\beta=3/4$, while the dashed horizontal lines indicate the region in which continuum luminosity and microlensing sizes are consistent, $\log(r_{L}/r_{\mu})=0\pm0.08\,\text{dex}$.  Interpreted as constraints on $\beta$, this implies $0.37 < \beta < 0.43$ (68\% confidence). This value of $\beta \approx 0.4$ is significantly shallower than the thin disk value of $\beta=0.75$.   


We have so far approximated the inner disk edge at $r_{\text{in}}=0$, but a finite inner edge will partially reconcile the size inconsistency.  To quantify the impact of the inner edge, we modify the temperature profile of Equation~\ref{eqn:t_v_beta} as
\begin{equation}
T(r) \propto r^{-\beta}\left( 1 - \sqrt{\frac {r_{\text{in}}}{r}} \right)^{(1/4)}.
\label{eqn:t_of_r}
\end{equation}
This is holds exactly for the thin disk model, though it only captures the approximate behavior for general values of $\beta$.  
Based on the disk sizes and black hole masses in Table~\ref{tab:qso_stats}, we choose to fix the inner edge at a typical value of $r_{\text{in}}=0.1r_{\lambda, 2500\text{\AA}}$.  This is roughly at or immediately outside of the innermost stable circular orbit (ISCO) of the central black hole, conservatively assuming a non-spinning (Schwarzchild) black hole.
Under this modification, we find the curve indicated by the dotted line in Figure~\ref{fig:tslope}.  For high values of $\beta$ the inner edge significantly alters the expected ratio.  In the gray region constrained by the data, however, the addition of the inner edge makes a negligible change in our inferred $\beta$.


Although microlensing sizes do not change with disk shape, they can be impacted by contaminating flux from large scales. Emission lines in the broad-line region, for example, account for significant quasar flux.  Even if a line center does not fall within our observing band, wings of the lines can contribute flux on the order of $30\%$ at $\lambda_{\text{rest}}=2500\,\text{\AA}$
\citep[e.g.,][]{maoz1993a, slus2007a}. 
This effect has been analyzed in previous multi-epoch microlensing studies \citep[e.g.,][]{dai2010a} and is shown to reduce the measured microlensing size $r_{\mu}$ by 20-50\% for 30\% contamination.  This shows that the measured microlensing size decreases roughly in proportion to the broad-line region contamination, $r_{\mu} \propto 1-F_{\text{BLR}}/F_{\text{tot}}$.
Though we expect the contamination in our aggregated sample to be less than 30\%, we adopt this value as an approximate upper limit. 

Broad-line contamination will also reduce the estimated luminosity size $r_{L}$, although by a smaller percentage because our models scale as $r_{L} \propto F_{\lambda, \text{obs}}^{1/2}$, following Equation~\ref{eqn:rlambda_flux}.  For 30\% contamination this decreases $r_L$ by 16\%.  The scale factor $C(\beta)$ also changes because the continuum half light radius is no longer formally determined by the accretion disk alone, although this effect is smaller than size decrease due to lower flux.
We plot the combined effect of this contamination on $\log({r_{L}}/{r_{\mu}})$ as the dot-dashed line in Figure~\ref{fig:tslope}.  Broad-line contamination brings the inferred $\beta$ closer to the thin disk prediction, giving $0.44 < \beta < 0.52$.  While a more significant broad line contamination would bring the inferred $\beta$ even closer to the thin disk prediction, a contamination percentage significantly greater than 30\% seems physically unlikely.

Light emitted from the accretion disk may also be scattered back into the line of sight at larger scales \citep[e.g.][]{dai2010a}.  This can come from scattering off of the outer disk edge or torus or as polar scattering from the surrounding medium \citep{goos2007a}.
We approximated this scattered light as a constant additive term which will not be microlensed because of the large scattering scales.  This will affect the microlensing size in the same way as broad-line region contamination, decreasing $r_{\mu}$ proportionally to the scattered light percentage.  Because scattered light originates from the disk, however, the disk luminosity, and therefore luminosity size, $r_{L}$, are effectively unchanged, although we still adjust $C(\beta)$ as with broad line contamination.

The scattered light percentage can be constrained through polarization measurements.  For example, \citet{huts2015a} explore a scattered light percentage between 17--44\% in the lensed quasar H1413+117, though they do not favor any particular value.  The findings of \citet{goos2007a} suggest that quasars, with low typical inclination angles, will likely have less scattered light than Seyferts.  Based on these ranges, we adopt a typical scattered light percentage of  $30\%$ and show the effect of scattering on the temperature profile as the dashed line in Figure~\ref{fig:tslope}.  Scattering has the largest impact on the inferred $\beta$, allowing values as high as $0.48 < \beta < 0.56$, which gives $\beta < 0.75$ at ${\sim}4\,\sigma$ significance.  While the true scattered light percentage could exceed 30\%, there is at present no strong evidence of high scattering fractions.  In such a case, scattered light would need to compose 61\% of the total flux to bring our temperature slope into agreement with the thin disk prediction at $2\,\sigma$.

Accounting for this range of systematic shifts, we find an improved constraint on quasar accretion disk temperature profiles $0.37 < \beta < 0.56$.  We plot the temperature profile the central value $\beta=0.47$ in Figure~\ref{fig:temp_v_r}, including the steeper thin disk temperature profile for comparison. Even if we allow for the case of both 30\% scattered light and 30\% broad line region contamination we still find $\beta < 0.75$ at $3\,\sigma$ confidence.  

This range of temperature profile slopes could be produced, for example, by the phenomenological disk wind model of \citet{li2019a}.  In this model mass outflows from an underlying thin disk increases sizes measured by microlensing or reverberation mapping techniques.  This model is completely analogous to ours, in which a disk wind parameter $s$ gives an effective temperature profile identical to Equation~\ref{eqn:t_v_beta} with $\beta=(3-s)/4$.  Our estimate of $\beta\approx0.47$ translates to a disk wind parameter of $s\approx1.1$, consistent with their findings.  While the model of \citet{li2019a} does not arise directly from accretion disk physics, a similar temperature profile slope may be produced in other disk wind models \citep[e.g.][]{slon2012a, yuan2012a, laor2014a, tomb2015a}.   In particular, Figure 5 of \citet{laor2014a} demonstrates that their disk wind model flattens the effective temperature profile.



\subsection{Size Ratio Variation with Disk Atmospheres}

\begin{deluxetable}{lc}
\tablecaption{The weighted mean ratio of continuum sizes estimated from luminosity to those measured by microlensing, $\log(r_{L}/r_{\mu})$.  We show the change for the thin disk model, a $\beta$ disk model at $\beta=0.4$ (which corresponds to the disk wind model of \citet{li2019a} with $s=1.4$), and the \citet{hall2018a} constant atmosphere model under two different densities. \label{tab:rlum_model}}
\tablehead{
\colhead{Model} & \colhead{$\log(r_{L}/r_{\mu})$}
}
\startdata
Thin Disk & $ -0.57\pm 0.08$ \\
$\beta=0.4$ Disk & $ -0.02\pm 0.08$ \\
Constant Atmosphere Disk $\log(\rho_{\text{atm}}/\rho)=-3$ & $ -0.29\pm 0.08$ \\
Constant Atmosphere Disk $\log(\rho_{\text{atm}}/\rho)=-4$ & $ -0.10\pm 0.08$ \\
\enddata
\end{deluxetable}

Our analysis so far assumes that accretion disks are homogenous with optically thick blackbody emission varying with radius with a power-law form for the surface temperature.  However, none of these assumptions need necessarily hold.  For example, \citet{hall2014a} showed that shock heating could alter a disk's temperature profile and inferred size.  In another work \citet{jian2016a} found that an iron opacity bump changes disk structure at radii comparable to our measured half light radii.  This could manifest in a broken power law temperature profile.  \citet{hall2018a} demonstrated that with optically thin scattering atmospheres, disks can have a flatter observed spectral energy distribution than that of the fiducial optically thick disk.  Depending on the characteristics of the disk atmosphere, these non-blackbody spectra can produce luminosity size estimates consistent with microlensing size measurements.  These models actually produce a steeper surface temperature profile slope than the thin disk, but the spectral slope of the emergent flux is broadened by the non-blackbody atmosphere.

For a comparison to the non-blackbody atmosphere model of \citet{hall2018a}, we must numerically calculate $r_{L}$ from the definition of the half-light radius
\begin{equation}
\frac{L_{\nu, 1/2}}{L_{\nu, \text{rest}}} = \frac{\int_{r_{\text{in}}}^{r_{L}} 2\pi r\,F_{\nu}(r)dr}{\int_{r_{\text{in}}}^{\infty} 2\pi r\,F_{\nu}(r)dr} = \frac{1}{2}.
\label{eqn:rhalf}
\end{equation}
Here all frequencies are in the rest frame of the quasar.  This equation retains the assumption of constant flux in each annular ring, but allows a generalized flux spectrum.  For $F_{\nu}$ we use the constant atmosphere flux model, Equation~1 of \citet{hall2018a}, where the surface temperature $T_s(r)$ is found according to \citet{hall2018a} Section~2.4.  We show this numerical $T_s(r)$ for a typical atmospheric density of $\log(\rho_{\text{atm}}/\rho)=-4$ in Figure~\ref{fig:temp_v_r}.  As discussed in \citet{hall2018a}, the surface temperature approaches the thin disk temperature from above at high radii.  At small radii, however, deviations from blackbody emission lead to $T_s$ is much higher than the thin disk.

We enforce that the model reproduces the observed flux, $F_{\nu, \text{obs}}$, by
\begin{align}
\begin{split}
L_{\nu, \text{rest}} &= 4\pi D_{L}^2 F_{\nu, \text{obs}}  /(\chi(i)(1+z)) \\&= 2\int_{r_{\text{in}}}^{\infty} 2\pi r\,F_{\nu, \text{rest}}(r)dr.
\label{eqn:lum_nu}
\end{split}
\end{align}
Here $z$ is the quasar redshift, $D_{L}$ is the luminosity distance, and $i$ is the viewing inclination.  The factor $\chi(i)$ corrects for anisotropy in the emitted flux, which can be due to the geometry of the emitting region, limb darkening, or general relativistic warping near the inner disk edge \citep{krol1999a, misr2003a}.  For our model we assume a flat thin disk and neglect relativistic effects, which gives $\chi(i) = 2\cos(i)$ \citep{misr2003a}.  We fix the inclination angle to $60\degr$ for consistency with reported microlensing sizes.  As in \citet{morg2010a} we adopt a nominal Eddington fraction of $f_{\text{edd}} = L/L_{E} = 1/3$ and an efficiency of $\eta=0.06$ for each model.

From the observed fluxes in the 15 quasars, we calculate the estimated luminosity size $r_{L}$ for the thin disk, our $\beta$ model (which also holds for the \citet{li2019a} model), and the constant atmosphere disk model in \citet{hall2018a}.  Here we use the microlensing sizes $r_{\mu}$ (without considering any large-scale flux contamination).  For the \citet{hall2018a} model the outer disk atmospheric density at each radius is assumed to be a constant ratio of the central density.  We select two scale factors, $\log(\rho_{\text{atm}}/\rho)=-3$ and $\log(\rho_{\text{atm}}/\rho)=-4$, which are favored by \citet{hall2018a}.  We show the weighted mean ratio, expressed as $\log(r_{L}/r_{\mu})$, in  Table~\ref{tab:rlum_model}, where we see that our $\beta\approx0.4$ disk and the constant atmosphere model with $\log(\rho_{\text{atm}}/\rho)=-4$ yield consistent luminosity sizes.

\subsection{Continuum Size Variation with Black Hole Mass}
\label{sec:size_v_mbh}

\begin{figure}[t]
\centering
\includegraphics[width=0.5\textwidth]{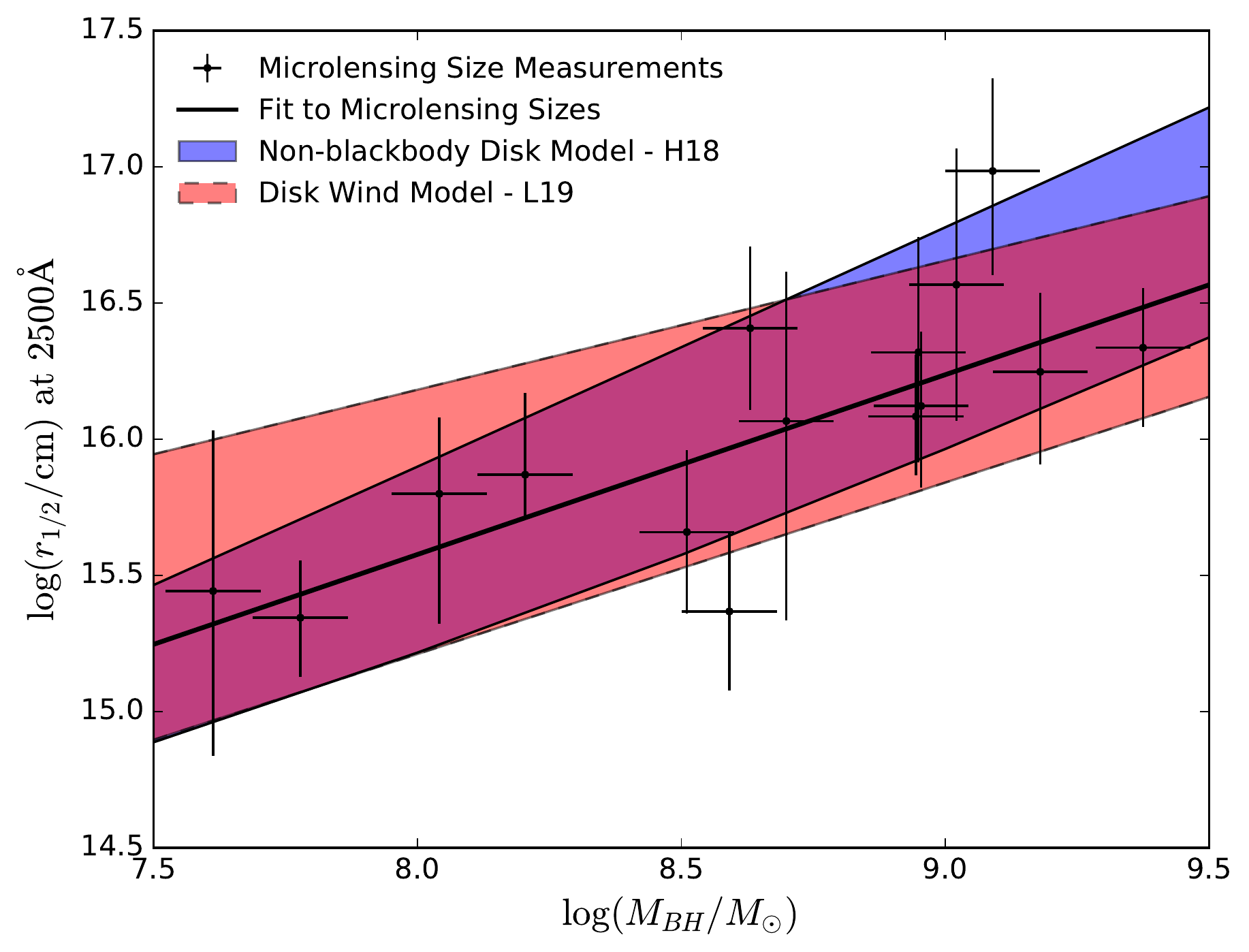} 
\caption{Quasar Accretion Disk Size-Black Hole Mass Relation with disk sizes at $\lambda_{\text{rest}} = 2500\text{\AA}$.  As in Figure~1 we show microlensing size measurements and best fit line, but now compare these measurements to predictions from two alternative disk models.  The light blue band (bound by solid lines) covers a range of constant atmosphere models following \citet{hall2018a} and ranging from $-1 \leq \log(\rho_{\text{atm}}/\rho) \leq -5$. In orange (bound by dashed lines) we show a range of disk wind models from \citet{li2019a} with the disk wind parameter between $0.3 \leq s \leq 1.1$.  The purple region is spanned by both models.}
\label{fig:r_v_mbh_models}
\end{figure}

As another means to distinguish models, we compare predictions for the continuum emission size as a function of black hole mass.  Here we use the constant atmosphere model with the range of atmospheric densities $-1 \leq \log(\rho_{\text{atm}}/\rho) \leq -5$.  For the disk wind model, we explore the range of disk wind parameters $0.3 \leq s \leq 1.1$.  Figure~\ref{fig:r_v_mbh_models} illustrates that under these ranges of atmospheric densities and disk wind parameters, both the models of \citet{hall2018a} and \citet{li2019a} can reproduce the empirical relationship.  

These two models do, however, predict different slopes of $r_{2500\text{\AA}}$ vs. $M_{\text{BH}}$.  The constant atmosphere model of \citet{hall2018a}, although not strictly a power law, gives an approximate slope of $0.74 < \log(r_{2500\text{\AA}})/\log(M_{\text{BH}}) < 0.88$.  The model of \citet{li2019a} predicts a shallower slope of $0.47 < \log(r_{2500\text{\AA}})/\log(M_{\text{BH}}) < 0.63$.  While both are consistent with the measured scaling $\log(r_{2500\text{\AA}})/\log(M_{\text{BH}})=0.66\pm0.15$, higher precision measurements may soon enable a distinction between these models.
 
\section{Discussion}
\label{sec:discussion}

We have found that the inconsistency between microlensing size measurements and luminosity size estimates of quasar UV continuum emission regions can be reconciled if the quasar accretion disk temperature profile is shallower than that of the thin disk.  The size difference cannot be completely explained by contamination from other emission sources nor by the impact of the inner disk edge.  We conclude that for a simple optically thick continuum emission region, the effective temperature profile must be much shallower than a standard thin disk to reproduce observed microlensing sizes. 

Our findings are in excellent agreement with those from spectral indexing, under models with little to no dust extinction, which measure $\beta\approx0.57$ \citep{davi2007a, gask2008a}.   Reverberation mapping \citep[e.g][]{edel2017a, edel2019a} generally measures slopes in agreement with the thin disk model, although \citet{jian2017a} found evidence for a shallower temperature profile.  Furthermore, most RM findings do not exclude the temperature profile we found here.

The shallower temperature profile we find here is, however, in tension with many chromatic microlensing studies.  Most chromatic microlensing measurements give a steeper-than-thin disk profile of $\beta \approx 1$ \citep{muno2011a, mott2012a, blac2015a, muno2016a} while some return a still steeper estimate of $\beta \approx 1.25$ or higher \citep{blac2011a, jime2014a, mott2017a}.  These measurements are inconsistent with our estimate of $\beta < 0.56$.  \citet{bate2018a} recently ran simulations which found that low chromatic variability in a lensed quasar system can lead to systematically steeper temperature profile slopes.  Indeed \citet{roja2014a} and \citet{bate2018a} analyzed several systems with high chromaticity and measured temperature profile slopes as low as $\beta = 0.43$.  However, the analysis of other high chromaticity lenses return values of $\beta \approx 0.9$, still steeper than the thin disk prediction \citep{muno2011a, mott2012a}.  \citet{bate2018a} point out that their simulations may not be completely general and it remains unclear whether a better understanding of chromaticity alone will be sufficient to reconcile chromatic microlensing findings with ours.

This inconsistency may also highlight the limitations of our own model dependent estimate of $\beta$.  If the underlying disk structure deviates strongly from the standard thin disk model, for example with large disk inhomogeneities \citep{dext2011a}, then our analysis may need significant modification.
We also have not accounted for the width of our filters or potential geometric effects from the nonzero thickness of the disk, although we expect these to have only a modest impact our inferred $\beta$.
Furthermore we have assumed that the size offset comes from a change in temperature profile slope, though a change in temperature profile normalization could also reconcile the size inconsistency.  A physical mechanism for such a normalization offset is, however, not immediately apparent and is difficult to produce under the standard thin disk model.  This would require, for example, a significant systematic offset in black hole masses.




From Section~\ref{sec:results}, it is clear that multiple models are consistent with our measurements, even including the constraint from black hole masses.  Our value of $\beta\approx0.47$ is also consistent with the temperature profile slope predicted by the slim disk models of \citet{abra1988a} and \citet{szus1996a}.   \citet{take2020a} presented a model of a rapidly accreting black hole which also predicts a scaling of $\beta=0.5$ (see Equation 26). Furthermore \citet{dext2011a} showed that disk inhomogeneities could also flatten the spectrum and explain the observed size difference, though the authors caution that this model is not necessarily physical.  Iron opacity could also lead to a broken power law temperature profile slope \citep{jian2016a}, although we presently lack sufficient observational constraints from microlensing alone to test this model. Comparison of these models to other quasar observations \citep[e.g.][]{davi2007a, shul2012a} will be necessary to fully confirm or refute any proposed models.

We have not considered here any of the recent numerical GRMHD models. We expect that these simulations will generally predict disks with non-blackbody spectra and inhomogeneities, likely including a departure from our assumption of constant annular flux.  Further publications of GRMHD disk azimuthally averaged surface temperature predictions, or better still, half-light continuum size predictions, along with size scaling with black hole mass, would enable ready comparison with our measurements.  \citet[][private communication]{jian2019a}, showed that their model readily produces values of $\beta\approx0.34$ above ${\sim}10$ gravitational radii, suggesting that this simulation qualitatively agrees with our findings, but further work is needed to produce a robust comparison.

Our results as given here are still model dependent, hinging on annular blackbody flux and a well-behaved disk surface temperature profile slope.  Our next step is to conduct multi-wavelength analyses of lensed quasars using the multi-epoch variability technique, expanding to H-band lightcurves which will measure the quasar continuum region size in the rest-frame optical.  While observationally expensive, this is the most bias-free method available to concretely measure disk temperature profiles in individual systems and will be independent of any accretion disk model. Fortunately the Large Synoptic Survey Telescope (LSST) \citep{lsst2009a} will provide many lightcurves suitable for such an analysis.

\acknowledgments

We thank Jason Dexter for helpful discussions on quasar accretion disk modeling.  We also thank Yan-Fei Jiang for generating the temperature profile in his recent simulation, providing estimates on the predicted temperature profile slope, and for helpful comments on this work.  We are also grateful to Patrick Hall for providing valuable feedback.

This material is based upon work supported by the National Science Foundation under grant AST-1614018 to M.A.C. and C.W.M.






\clearpage

\bibliographystyle{aasjournal}
\bibliography{usna_bibtex_archive}

\end{document}